\documentclass[aps,preprint,superscriptaddress,amsmath,amssymb,showpacs]{revtex4-2}
\usepackage{amssymb,bm}
\usepackage{graphicx}
\usepackage{amsmath}
\usepackage{color}
\usepackage[colorlinks=true,citecolor=blue,urlcolor=blue]{hyperref}
\usepackage{comment}

\begin{document}

\title{Manifestation of the electric dipole moment in the decays of $\tau$ leptons produced in $e^+e^-$ annihilation}
\author{I.V.Obraztsov}\email{ivanqwicliv2@gmail.com}
\author{A. I. Milstein}\email{A.I.Milstein@inp.nsk.su}

\affiliation{Budker Institute of Nuclear Physics, 630090 Novosibirsk, Russia}
\affiliation{\textit{Novosibirsk State University, 630090, Novosibirsk, Russia}}
\date{\today}

\begin{abstract}
$\mbox{CP}$-odd asymmetries in the processes  $e^+e^-\rightarrow \tau^+\pi^-\nu_\tau$, $e^+e^-\rightarrow \pi^+\tau^-\bar\nu_ \tau$, $e^+e^-\rightarrow \tau^+\rho^-\nu_\tau$, $e^+e^-\rightarrow \rho^+\tau^-\bar\nu_\tau $, $e^+e^-\to \tau^+e^-\nu_\tau{\bar\nu}_e\,,\,$ and $\,e^+e^-\to \tau^-e^+\nu_e{\bar\nu}_\tau$  are investigated with account for longitudinal polarization of electron (or positron) beam. These asymmetries is a manifestation of electric dipole form factor $F_3^\tau\equiv b$ in the $\gamma\tau^+\tau^-$ vertex. It is shown that, to measure $\mbox{Im}\,b$ in the specified processes, polarization is not needed, while to measure $\mbox{Re}\,b$ it is required. The processes $e^+e^-\to \pi^+\pi^-\nu_\tau{\bar\nu}_\tau$, $e^+e^-\to e^+e^-\nu_\tau{\bar\nu}_\tau\nu_e\bar\nu_e$,  $e^+e^-\to \mu^+\mu^-\nu_\tau{\bar\nu}_\tau\nu_\mu\bar\nu_\mu$, $e^+e^-\to \mu^+e^-\nu_\tau{\bar\nu}_\tau\nu_\mu\bar\nu_e$, and $e^+e^-\to \mu^-e^+\nu_\tau{\bar\nu}_\tau\nu_e\bar\nu_\mu$ are also discussed for the case of unpolarized electron and positron beams. In the latter cases it is possible to measure $\mbox{Re}\,b$ using the differential cross section over momenta of both registered particles.
\end{abstract}

\maketitle

\section{Introduction}
One of the ways to search for New Physics is
precision measurement of the electric dipole moment $d_l$ of a charged lepton ($l=e,\mu,\tau$). The  value of  $d_l$ predicted by the Standard Model (SM) is too small for experimental measurement. Therefore, the observation of electric dipole moment or its manifestation would directly demonstrate the existence of New Physics.

The manifestation of the lepton electric dipole moment can be sought in the process of $l\bar l$ pair production in $e^+e^-$ annihilation. The general form of  $\gamma l \bar l$ vertex can be represented as
\begin{align}\label{genvert}
	&\Gamma^\mu=-ie\Bigg\{ F_1^l(k^2)\gamma^\mu +\dfrac{\sigma^{\mu\nu}k_\nu}{2m_l} \Big [iF_2^l(k^2) +F_3^l(k^2)\gamma_5 \Big ] +\Big( \gamma^\mu - \dfrac{2k^\mu \, m_l}{k^2}\Big)\gamma_5 F_4^l(k^2) \Bigg\}\,,
\end{align}
where $m_l$ is the lepton mass, $e<0$ is the electron charge, $k$ is the 4-momentum of a photon, $F_1^l(k^2)$ is the Dirac form factor, $F_2^l(k ^2)$ is the Pauli form factor, $F_3^l(k^2)$ is the electric dipole form factor, $F_4^l(k^2)$ is the anapole form factor, $\sigma^{\mu\nu} =i/2[\gamma^\mu,\gamma^\nu]$, $\gamma_5=-i\gamma^0\gamma^1\gamma^2\gamma^3$. In the limit $k^2 \to 0$ these form factors are
\begin{align}
	& F_1^l(0)=1\, ,\quad F_2^l(0)=\mu'_l\,\dfrac{2m_l}{e}\, ,\quad F_3^l(0)=d_l\,\dfrac{2m_l}{e}\,,\quad
	F_4^l(0)=0\,,
\end{align}
where $\mu'_l$ is the anomalous magnetic moment. It follows from \eqref{genvert} that a violation of $\mbox{P}$ and $\mbox{T}$ parities leads to the appearance of $F_3^l(k^2)$, while $F_4^l(k^2)$ is related to  violation of $\mbox{P}$-parity alone. Assuming that the $\mbox{CPT}$ theorem holds, violation of $\mbox{T}$-parity is equivalent to violation of $\mbox{CP}$-parity. Thus, $d_l$ occurs due to $\mbox{CP}$ violations.

For all charged leptons, predictions of  $\mu'_l$  in  SM \cite{Aoyama:2019ryr,Volkov:2019phy,Aoyama:2020ynm,Eidelman:2007sb}  can be experimentally verified \cite{Fan:2022eto,Muong-2:2021ojo,DELPHI:2003nah}. For $d_l$ the situation is essentially different. An estimate of $d_l$ in  SM \cite{Booth:1993af,Mahanta:1996er,Yamaguchi:2020eub,Yamaguchi:2020dsy}  gives $|F_3^e(0)|<|F_3^\mu(0)|< |F_3^\tau(0)|\approx 10^{-23}\ll 1$. The sensitivity of modern experiments does not allow one to  measure  $F_3^\tau(0)$ with an accuracy of $10^{-23}$. Therefore, extracting a non-zero value of $F_3^\tau(0)$ from the experiment would be a discovery of New Physics.  In Refs.~ \cite{Barr:1988mc,Grifols:1990ha,Escribano:1993pq,Escribano:1996wp,Taylor:1997tn,OPAL:1998dsa,L3:1998gov,DELPHI:2003nah,L3:2004ful,Blinov:2008mu,Grozin:2008nw} upper limits were set to $|F_3^\tau(k^2)|$, and in Refs.~\cite{ARGUS:2000riz,Belle:2002nla,Belle:2021ybo} upper limits were set to $\mbox{Re} \, F_3^\tau(k^2)$ and $\mbox{Im}\, F_3^\tau(k^2)$ separately.

In our work,  the processes $e^+e^-\rightarrow \tau^+\pi^-\nu_\tau$, 
 $e^+e^-\rightarrow \pi^+\tau^-\bar\nu_\tau$, $e^+e^-\rightarrow \tau^+\rho^-\nu_\tau$, $e^+e^-\rightarrow \rho^+\tau^-\bar\nu_\tau$, $e^+e^-\to \tau^+e^-\nu_\tau{\bar\nu}_e\,,\,$ and $\,e^+e^-\to \tau^-e^+\nu_e{\bar\nu}_\tau$ are studied for longitudinally polarized electron beam. The   processes  $e^+e^-\rightarrow \pi^+\pi^-\nu_\tau\bar\nu_\tau$, $e^+e^-\to e^+e^-\nu_\tau{\bar\nu}_\tau\nu_e\bar\nu_e$,  $e^+e^-\to \mu^+\mu^-\nu_\tau{\bar\nu}_\tau\nu_\mu\bar\nu_\mu$, $e^+e^-\to \mu^+e^-\nu_\tau{\bar\nu}_\tau\nu_\mu\bar\nu_e$, and $e^+e^-\to \mu^-e^+\nu_\tau{\bar\nu}_\tau\nu_e\bar\nu_\mu$ are discussed for unpolarized electron and positron beams. The asymmetric with respect to $\mbox{CP} $ transformation parts of the corresponding cross sections  are obtained for $e^+e^-$ invariant masses  $\sqrt{s}\ll m_Z$.  To derive these results, it is sufficiently to consider   $\gamma\tau^+\tau^-$ vertex  in the form
\begin{align}\label{tauvert}
	&\Gamma^\mu=-ie\Bigg [ \gamma^\mu +\dfrac{\sigma^{\mu\nu}k_\nu}{2M} F_3^\tau(k^2)\gamma_5\Bigg ]\,,
\end{align}
where $M$ is the $\tau$ lepton mass and $k^2=s$.

Measurement of $\mbox{CP}$ odd parts of the cross sections  can diminish the upper limits of both $\mbox{Re}\,F_3^\tau(s)$ and $\mbox{Im}\,F_3^\tau(s)$. Moreover,  polarization of beams simplifies measurement of $\mbox{Re}\,F_3^\tau(s)$. At present, there are no corresponding experiments with    longitudinally polarized electrons. However, such  experiments are planned to be carried out at the Super-Charm-Tau factory (SCTF) \cite{TauFactory}. This  collider, having high luminosity $\sim 10^{35}\,\mbox{cm}^{-2}\,\mbox{s}^{-1}$ and $\sqrt{s}$ from  
$3$ to $5-7\,\mbox{GeV}$,  will become an intense source of  $\tau$ leptons. Note that
$\mbox{Im}\,F_3^\tau(0)=0$ due to the CPT theorem. If one assume that a typical size of New Physics $\Lambda_{NP}\gg M$, then one should expect that $\mbox{Im}\,F_3^\tau(s)\ll \mbox{Re}\,F_3^\tau(s)$ at $s\gtrsim M^2$. This why the use of longitudinal polarization of electrons to study the dipole moment of $\tau$ lepton is very important.

\section{$e^+e^-\to \tau^+\tau^-$}
To study the effect of polarization, let us consider a longitudinally polarized electron beam and an unpolarized positron beam. Since $\sqrt{s}\ll m_Z$, we neglect the contribution of the Z boson. Then the cross section $d\sigma_0$ of the process $e^+e^-\to \tau^+\tau^-$ in the center-of-mass frame is
\begin{align}
	&d\sigma_0=\dfrac{\beta\alpha^2}{4s}\,\left |\phi_\tau^\dagger\,H\,\chi_\tau\right|^2\,d\Omega_{\bm q}\,,\nonumber\\
	&H=\bm\sigma\cdot\bm e_\lambda-\dfrac{(\bm\sigma\cdot\bm q)(\bm e_\lambda\cdot\bm q)}{E(E+M)}+i\dfrac{b}{M}\,(\bm e_\lambda\cdot\bm q)\,,\nonumber\\
	&b=F_3^\tau(s)\,,\quad s=4E^2\,,\quad \beta=q/E\,,\quad \bm e_\lambda=\frac{1}{\sqrt{2}}(\bm e_x+i\lambda\bm e_y)\,.
\end{align}
Here $\alpha$ is the fine structure constant, $E$ is the electron energy, $\bm q$ is the momentum of the $\tau^-$ lepton, the vector $\bm e_z$ is directed along the electron momentum, $\lambda$ is the electron helicity, $\phi_\tau$ and $\chi_\tau$ are two-component spinors entering, respectively, into the positive-frequency and negative-frequency solutions of the Dirac equation for a $\tau$ lepton,
\begin{align}\label{wfin}
	&U_{\bf q}=\sqrt{\dfrac{E_q+M}{2E_q}}
	\begin{pmatrix}
	\phi_\tau\\
	\dfrac{\bm\sigma\cdot\bm q}{E_q+M}\,\phi_\tau\end{pmatrix}\, ,\quad 
V_{\bf -q}=\sqrt{\dfrac{E_q+M}{2E_q}}
\begin{pmatrix}
\dfrac{-\bm\sigma\cdot\bm q}{E_q+M}\,\chi_\tau\\
\chi_\tau\end{pmatrix}\,,
	\end{align}
where $E_q=\sqrt{\bm q^2+M^2}$.
Using the relation
$$e_\lambda^i e_\lambda^{*j}=\dfrac{1}{2}(\delta^{ij}-\Lambda^i\Lambda^j-i\epsilon^{ijk}\Lambda^k)\,,\quad \bm\Lambda =\lambda\bm e_z\,,$$
we obtain the cross section $d\sigma_0$ summed over polarizations of $\tau^+$ 
\begin{align}
	&d\sigma_0=\dfrac{\beta\alpha^2}{4s}\,\left[ 1-\dfrac{ q_\perp^2}{2E^2}+\bm\zeta\cdot\bm Z\right]\,d\Omega_{\bm q}\,,\nonumber\\
	&\bm Z=\mbox{Im}\,b\,\dfrac{q_\perp^2\,\bm q}{ME(E+M)}-\mbox{Im}\,b\,\dfrac{\bm q_\perp}{M}-\,\mbox{Re}\,b\,\dfrac{[\bm q_\perp\times\bm \Lambda]}{M}\nonumber\\
	&+\dfrac{M}{E}\bm \Lambda+\dfrac{(\bm q\cdot\bm \Lambda)\,\bm q}{E(E+M)}\,,
\end{align}
where $\bm q_\perp=\bm q- \bm \Lambda (\bm q \cdot \bm \Lambda)$, $\bm \zeta$ is the spin of $\tau^-$, and terms quadratic in $b$ are omitted. It is seen that the  linear in $b$ terms contribute only to the $\bm \zeta$-dependent part of the cross section. Study of various $\tau$ decay channels is a way to measure the polarization, which in turn makes it possible to measure $b$. The total cross section summed over the $\tau^-$ polarization is
 \begin{align}\label{sig0}
 	&\sigma_0=\dfrac{\pi\beta\alpha^2}{3E^2}\,\left( 1+\dfrac{M^2}{2E^2}\right)\,.
 \end{align}

\section{$e^+e^-\to \tau^+\pi^-\nu_\tau$ , $e^+e^-\to \tau^-\pi^+{\bar\nu}_\tau$}
Consider the cross section of the process $e^+e^-\to \tau^+\tau^-$ followed by the decay  $\tau^-\to \pi^-\nu_\tau$. Taking into account the smallness of the $\tau$  lepton  width,  $\Gamma_\tau\approx 2.27\,\mbox{meV}$ \cite{Workman:2022ynf},  we can make the substitutions
\begin{align}
&\dfrac{1}{{\hat q}-M}\rightarrow  \,\dfrac{2E_q}{{\cal E}^2-E_q^2+i\Gamma_\tau M}\,\sum_{\mu}U_{\bm q,\mu}\bar U_{\bm q,\mu} \,,\nonumber\\
&\dfrac{4E_q^2}{({\cal E}^2-E_q^2)^2+\Gamma_\tau^2 M^2}\rightarrow \dfrac{2\pi E_q}{M\Gamma_\tau}\,\delta({\cal E}-E_q)\,,
\end{align}
where ${\cal E}=q^0$. After that, the cross section of the process $e^+e^-\to \tau^+\pi^-\nu_\tau$ can be represented as
\begin{align}
&d\sigma_\pi^{(-)}(\bm k)=B_\pi\dfrac{\beta\alpha^2(E+M)\,d\Omega_{\bm q}\,d\bm k}{4\pi sM^2\,\omega_k}\,R^{(-)}\,\delta(E-\omega_k-|\bm q-\bm k|)\,,\nonumber\\
&R^{(-)}=\left |\phi^+_\nu\left[1+\dfrac{\bm\sigma\cdot\bm q}{E+M}\right]\,H\,\chi_\tau\right|^2\,,
\end{align}
where $B_\pi\approx 10.8\%$ \cite{Workman:2022ynf} is the branching ratio of $\tau\to \pi\nu$  decay, $\omega_k=\sqrt{\bm k^2+m_\pi^2 }$ and $\bm k$ are the pion energy and momentum, respectively, $\phi_\nu$ is a two-component spinor in the Dirac  spinor $U_{\bm Q}$ for neutrino, $\bm Q=\bm q-\bm k$ and $(\bm \sigma\cdot\bm Q)\phi_\nu=-Q\phi_\nu$.
We also neglected the pion mass $m_\pi$ compared to $M$ and took into account that the  matrix element of the $\tau\to \pi\nu$ is proportional to $\bar U_{\bm Q}\,U_{\bm q }$ \cite{Okun:1982ap}. After summing over polarizations $\chi_\tau$, we find
\begin{align}
&R^{(-)}=A_0-\bm n_{\bm Q}\bm A\,,\quad \bm n_{\bm Q}=\bm Q/Q\,,\nonumber\\
&A_0=\dfrac{2E}{E+M}\left[ 1-\dfrac{ q_\perp^2}{2E^2}+\dfrac{(\bm q\cdot\bm \Lambda)}{E}- \mbox{Im}\,b\,\dfrac{q_\perp^2}{E^2}\right]\,,\nonumber\\
&\bm A=\dfrac{2}{(E+M)}\Bigg\{\dfrac{M^2}{E}\bm \Lambda+\left[1+\dfrac{(\bm q\cdot\bm \Lambda)}{E}-\dfrac{ q_\perp^2}{2E^2}\right]\,\bm q\nonumber\\
&-\mbox{Im}\,b\,\bm q_\perp-\mbox{Re}\,b\,[\bm q\times\bm \Lambda]\Bigg\}\,.
\end{align}
Integrating over the angles of vector $\bm q$, we obtain
\begin{align}\label{sm}
	&d\sigma_\pi^{(-)}(\bm k)=B_\pi\dfrac{\alpha^2}{4E^3}\,\Big[\xi_0+\xi_1\cos\theta+\xi_2(3\cos^2\theta-1)\Big]\,d\omega_k\,d\Omega_{\bm k}\,,\nonumber\\
	&\xi_0=\dfrac{1}{3}\left(1+\dfrac{M^2}{2E^2}\right)+\dfrac{\mbox{Im}\,b}{3}\left(1-\dfrac{2\omega_k}{E}\right)\,,\nonumber\\
	&\xi_1=\dfrac{M^2}{4E\omega_k}-\dfrac{1}{2}\left(1-\dfrac{2\omega_k}{E}\right)\,,\nonumber\\
	&\xi_2=\dfrac{M^2(M^2-4E\omega_k)}{32E^2\omega_k^2}+\dfrac{1}{12}\left(1+\dfrac{M^2}{2E^2}\right)+\dfrac{\mbox{Im}\,b}{3}\left(1-\dfrac{\omega_k}{2E}-\dfrac{3M^2}{8E\omega_k}\right)\,.
\end{align}
Here $\cos\theta=\bm k\cdot\bm\Lambda/k=\lambda k_z/k$, the available pion energy range is determined by the relation $|2\omega_k-E|\leq q$. It is seen that the only term, which contains the helicity $\lambda$, is that proportional to $\xi_1$. However, this term is independent of $b$. Moreover, the cross section $d\sigma_\pi^{(-)}(\bm k)$ in \eqref{sm} contains only the imaginary part of $b$, which does not require a nonzero electron polarization.
For the cross section $d\sigma_\pi^{(+)}(\bm k)$ of the process $e^+e^-\to \tau^-\pi^+{\bar\nu}_\tau$, we similarly find
\begin{align}\label{sp}
	&d\sigma_\pi^{(+)}(\bm k)=B_\pi\dfrac{\alpha^2}{4E^3}\,\Big[\xi'_0+\xi'_1\cos\theta+\xi'_2(3\cos^2\theta-1)\Big]\,d\omega_k\,d\Omega_{\bm k}\,,\nonumber\\
	&\xi'_0=\dfrac{1}{3}\left(1+\dfrac{M^2}{2E^2}\right)-\dfrac{\mbox{Im}\,b}{3}\left(1-\dfrac{2\omega_k}{E}\right)\,,\nonumber\\
	&\xi'_1=-\xi_1=-\dfrac{M^2}{4E\omega_k}+\dfrac{1}{2}\left(1-\dfrac{2\omega_k}{E}\right)\,,\nonumber\\
	&\xi'_2=\dfrac{M^2(M^2-4E\omega_k)}{32E^2\omega_k^2}+\dfrac{1}{12}\left(1+\dfrac{M^2}{2E^2}\right)-\dfrac{\mbox{Im}\,b}{3}\left(1-\dfrac{\omega_k}{2E}-\dfrac{3M^2}{8E\omega_k}\right)\,.
\end{align}
We define the differential asymmetry $dA_\pi$ as follows
\begin{align}\label{api}
	&dA_\pi=\dfrac{d\sigma_\pi^{(-)}(\bm k)-d\sigma_\pi^{(+)}(-\bm k)}{2\sigma_0}\,,
\end{align}
where $\sigma_0$ is defined in \eqref{sig0}. We have
\begin{align}\label{da}
	&dA_\pi=\dfrac{B_\pi{\mbox{Im}\,b}\,d\omega_k\,d\Omega_{\bm k}}{4\pi q(1+M^2/2E^2)}\,\left[1-\dfrac{2\omega_k}{E}+(3\cos^2\theta-1)\left(1-\dfrac{\omega_k}{2E}-\dfrac{3M^2}{8E\omega_k}\right)\right]\,.
\end{align}
Since the asymmetry is proportional to $\mbox{Im}\,b$, its measurement is very important.

After integration over $d\Omega_{\bm k}$, we obtain
\begin{align}\label{da1}
	&dA_\pi=\dfrac{{B_\pi\mbox{Im}\,b}\,d\omega_k}{ q(1+M^2/2E^2)}\,\left(1-\dfrac{2\omega_k}{E}\right)\,.
\end{align}
As it should be, after integration over the pion energy, the asymmetry vanishes. Therefore, we define the total asymmetry $A_\pi$  as $dA_\pi$ \eqref{da1} integrated over $\omega_k$ from $(E-q)/2$ to $E/2$ (half of the allowed energy range),
\begin{align}\label{apiE}
	&A_\pi=\dfrac{B_\pi\,\mbox{Im}\,b}{ 4(1+M^2/2E^2)}\,\sqrt{1-\dfrac{M^2}{E^2}}\,.
\end{align}
Taking  in Eq.\eqref{da} the integral over $\omega_k$ in the region $(E-q)/2\leq \omega_k\leq (E+q)/2$, we obtain the angular asymmetry  
\begin{align}\label{daangle}
	&dA_{\pi}=-\dfrac{3\,B_{\pi}\,\mbox{Im}\,b\,d\Omega_{\bm k}}{ 16\pi (1+M^2/2E^2)}\,
	(3\cos^2\theta-1)
	\left[\dfrac{M^2}{2Eq}\ln\left(\dfrac{E+q}{E-q}\right)-1\right]\,.
\end{align}

\section{$e^+e^-\to \tau^+\rho^-\nu_\tau$ , $e^+e^-\to \tau^-\rho^+{\bar\nu}_\tau$}
To measure $\mbox{Re}\,b$, consider the decay of one $\tau$ lepton into $\rho$ meson with the momentum
$\bm p$, energy $\varepsilon_p=\sqrt{\bm p^2+m_\rho^2}$ and 4-polarization vector $f=(f_0,\,\bm f)$, where $m_\rho $ is the mass of $\rho$ meson. We define the differential asymmetry $dA_\rho$ as
\begin{align}\label{arho}
	&dA_\rho=\dfrac{d\sigma_\rho^{(-)}(\bm p,\,\bm f)-d\sigma_\rho^{(+)}(-\bm p,\,-\bm f)}{2\sigma_0}\,,
\end{align}
where $d\sigma_\rho^{(-)}(\bm p,\,\bm f)$ and $d\sigma_\rho^{(+)}(-\bm p,\,-\bm f)$ are the cross sections of the processes $e^+e^-\to \tau^+\rho^-\nu_\tau $ and  $e^+e^-\to \tau^-\rho^+{ \bar\nu}_\tau$, respectively. Using the  matrix element of decay $\tau\to \rho\nu$ \cite{Okun:1982ap},  we obtain as a result of straightforward calculations
\begin{align}\label{darhopol}
	&dA_\rho=\dfrac{3B_\rho\,d\varepsilon_p\,d\Omega_{\bm p}\,\,[C_1\,\mbox{Re}\,b+C_2\,\mbox{Im}\,b]}{2\pi pM^2(1+M^2/2E^2)(2+M^2/m_\rho^2)(1-m^2_\rho/M^2)^2}\,,\nonumber\\
	&C_1=\left[\dfrac{q}{p}\varepsilon_p\,P_2(x_0)-E P_1(x_0)\right]\,([\bm\Lambda\times\bm f]\cdot\bm p)\,f_0\,,\nonumber\\ &C_2=\dfrac{qp}{3}\left[\left(2+\dfrac{\varepsilon_p}{E}\right)f_0^2+\left(1-\dfrac{\varepsilon_p}{E}\right)\bm f^2-(\bm\Lambda\cdot\bm f)^2\right]\nonumber\\
	&+P_1(x_0)\,\Big[\dfrac{1}{2}[p^2-(\bm\Lambda \cdot\bm p)^2](\bm f^2-f_0^2) \nonumber\\
	&+Ef_0[(\bm\Lambda\cdot\bm p)(\bm\Lambda\cdot\bm f)-(\bm p\cdot\bm f)]
	+\dfrac{2q^2}{5E}f_0[(\bm\Lambda\cdot\bm p)(\bm\Lambda\cdot\bm f)-2(\bm p\cdot\bm f)]\Big]\nonumber\\
	&+\dfrac{q}{p}\,P_2(x_0)\,\Bigg\{\left(\dfrac{\varepsilon_p}{2E}(\bm f^2-f_0^2)-f_0^2\right)\left[(\bm\Lambda\cdot\bm p)^2-\dfrac{p^2}{3}\right]+\left[(\bm f\cdot\bm p)^2-\dfrac{p^2\bm f^2}{3}\right]\nonumber\\
	&-(\bm f\cdot\bm\Lambda)\left[(\bm\Lambda\cdot\bm p)(\bm f\cdot\bm p)-\dfrac{p^2}{3}(\bm f\cdot\bm\Lambda)\right]\Bigg\}\nonumber\\
	&+P_3(x_0)\,\dfrac{q^2}{Ep^2}f_0\,\Big\{(\bm f\cdot\bm p)(\bm\Lambda\cdot\bm p)^2-\dfrac{p^2}{5}[(\bm f\cdot\bm p)+2(\bm f\cdot\bm\Lambda)(\bm p\cdot\bm\Lambda)]\Big\}\,,\nonumber\\
	&x_0=\dfrac{2E\varepsilon_p-M^2-m_\rho^2}{2qp}\,,
\end{align}
where $B_\rho\approx25.5\%$ \cite{Workman:2022ynf} is the branching ratio of $\tau\to\rho\nu$ decay and $P_n(x)$ are Legendre polynomials. Thus, for a polarized electron beam and  a polarized $\rho$ meson, the contribution of  $\mbox{Re}\,b$ does not vanish. Note that
\begin{align}
	& dA_\rho\Big|_{\lambda=+1}-dA_\rho\Big|_{\lambda=-1}=\dfrac{3B_\rho\,\mbox{Re}\,b\,C_1\,d\varepsilon_p\,d\Omega_{\bm p}}{\pi pM^2(1+M^2/2E^2)(2+M^2/m_\rho^2)(1-m^2_\rho/M^2)^2}\,.
\end{align}
Then we perform summation over polarizations of $\rho$ meson, using the formula
$$\sum_{pol} f^\mu f^\nu= -g^{\mu\nu}+\dfrac{p^\mu p^\nu}{m_\rho^2}\,,$$
and obtain 
\begin{align}\label{darho1}
		&\sum_{pol}C_1=0\,,\quad \sum_{pol}C_2=C_{21}\,F+C_{22}\,,\nonumber\\ 
		&C_{21}=-\dfrac{qp}{3}+P_1(x_0)\,\left(-\dfrac{3}{2}m_\rho^2+\dfrac{7}{5}E\varepsilon_p
		-\dfrac{2\varepsilon_p}{5E}M^2\right)\nonumber\\
	&+\dfrac{q}{p}P_2(x_0)\left(\dfrac{3\varepsilon_p}{2E}m_\rho^2-\dfrac{5}{3}\varepsilon_p^2
	+\dfrac{2}{3}m_\rho^2\right)+P_3(x_0)\,\dfrac{3\varepsilon_pq^2}{5E}\,,\nonumber\\
&C_{22}=\dfrac{p(M^2-2m_\rho^2)[m_\rho^2 E+M^2(E-2\varepsilon_p)]}{6m_\rho^2E\,q}\,,\nonumber\\
&F=\dfrac{1}{3m_\rho^2}\left[3(\bm\Lambda\cdot\bm p)^2-	p^2\right]\,.
\end{align}
In this formula, the contribution $\propto\mbox{Re}\,b$ is absent even for the case of polarized electrons. After integration over the angles of the vector $\bm p$, the term  $\propto F$ vanishes, and  the asymmetry reads
\begin{align}\label{darhoeps1}
	&dA_\rho=\dfrac{6B_\rho\,d\varepsilon_p\,C_{22}\,\mbox{Im}\,b}{ pM^2(1+M^2/2E^2)(2+M^2/m_\rho^2)(1-m^2_\rho/M^2)^2}\,.
	\end{align}
The allowed region of the energy $\varepsilon_p$ is given by the relation
$$\left|2\varepsilon_p-\left(1+\dfrac{m_\rho^2}{M^2}\right)E\right|\leq   q\left(1-\dfrac{m_\rho^2}{M^2}\right)\,.      $$
After integration over $\varepsilon_p$ in Eq.~\eqref{darhoeps1}, the asymmetry vanishes. Therefore, we define the total asymmetry  $A_\rho$ as a result of integration of Eq.~ \eqref{darhoeps1}  over $\varepsilon_p$ in the region
$$(1+m_\rho^2/M^2)E-(1-m_\rho^2/M^2)q<2\varepsilon_p< (1+m_\rho^2/M^2)E\,.$$
One has
\begin{align}\label{arhoE}
	&A_\rho=\dfrac{B_\rho\,\mbox{Im}\,b}{ 2(2+M^2/E^2)}\,\sqrt{1-\dfrac{M^2}{E^2}}\Bigg(\dfrac{M^2-2m_\rho^2}{M^2+2m_\rho^2}\Bigg)\,.
\end{align}
Note that
 \begin{align}\label{arhoE1}
 	&A_\rho=\dfrac{B_\rho}{ B_\pi}\Bigg(\dfrac{M^2-2m_\rho^2}{M^2+2m_\rho^2}\Bigg)\,A_\pi\,.
 \end{align}

To determine  the polarization of  $\rho$ meson, it is possible to measure the main decay channel of  $\rho$ meson with momentum $\bm p$ into two pions with momenta $\bm k_1$ and $\bm k_2$. The corresponding asymmetry can be obtained from Eq.~
\eqref{darhopol} by the obvious substitution
\begin{align}\label{darhopipi}
	&\frac{d\bm p}{2\varepsilon_p(2\pi)^3}\,f^\mu\,f^\nu\rightarrow  \,\dfrac{f^2_{\rho\pi\pi}\,d\bm k_1\,d\bm k_2\,(k_1-k_2)^\mu(k_1-k_2)^\nu}{ 4\omega_{1}\omega_{2}(2\pi)^6 [(s_1-m_\rho^2)^2+\Gamma_\rho^2m_\rho^2]}\,.
\end{align}
Here $s_1=(k_1+k_2)^2$, $\Gamma_\rho=149.1\,\mbox{MeV}$ \cite{Workman:2022ynf} is the $\rho$ meson width, $\omega_1=|\bm k_1|$, $\omega_2=|\bm k_2|$, and the constant $f^2_{\rho\pi\pi}$ is
$$ f^2_{\rho\pi\pi}=\dfrac{48\pi\Gamma_\rho}{m_\rho\,(1-4m_\pi^2/m_\rho^2)^{3/ 2}}\,.$$

\section{$e^+e^-\to \tau^+e^-\nu_\tau{\bar\nu}_e\,,\,$ $\,e^+e^-\to  \tau^-e^+\nu_e{\bar\nu}_\tau$ }
\begin{figure}[h]
	\centering
	\includegraphics[width=0.55\linewidth]{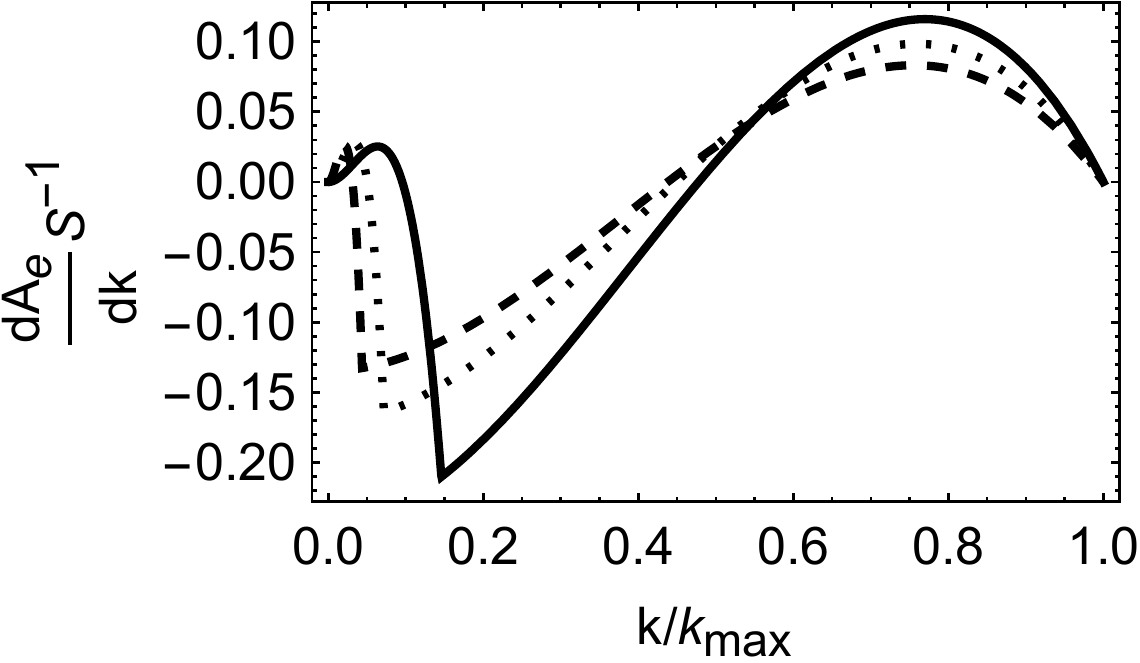}
	\caption{The asymmetry $dA_e/dk$ in units of $S=B_e\mbox{Im}\,b/M$ as a function of $k/k_{max}$ for a few values of $E$, $k_{max}=(E+q)/2$. Solid curve: $E=1.5\,M$, dotted curve: $E=2\,M$, dashed curve: $E=2.5\,M$.}
	\label{asek}
\end{figure}

Let us consider the cross sections $d\sigma_e^{(-)}(\bm k)$ and $d\sigma_e^{(+)}(\bm k)$  of the processes $e^+e^-\to \tau^+e^-\nu_\tau{\bar\nu}_e $ and $\,e^+e^-\to \tau^+\tau^-\to \tau^-e^+\nu_e{\bar\nu}_\tau$, respectively, where $\bm k$ is the electron (positron) momentum.  We define the asymmetry $ dA_{e}$ as
\begin{align}\label{dae}
	&dA_{e}=\dfrac{d\sigma_{e}^{(-)}(\bm k)-d\sigma_{e}^{(+)}(-\bm k)}{2\sigma_0}\,.
\end{align}
Then we use the  matrix element of decay $\tau^-\to e^-\nu_\tau{\bar\nu}_e\,$ \cite{Okun:1982ap} and perform the integration of cross sections over the neutrino and antineutrino momenta. We have 
\begin{align}\label{dae1}
	&dA_{e}=\dfrac{6B_{e}\,d\Omega_q\,d\bm k}{ (2\pi)^2M^6(1+M^2/2E^2)\,k}\,
	[4(kE-\bm k\cdot\bm q)-M^2]\nonumber\\
	&\times\,\Big[\mbox{Re}\,b\,[\bm k\times\bm q]\cdot\bm\Lambda+
	\mbox{Im}\,b\,(\bm k\cdot\bm q_\perp-\dfrac{k}{E}\,q_\perp^2)\Big]\,,
\end{align}
where $B_e\approx 18\%$ \cite{Workman:2022ynf} is the branching ratio of $\tau^-\to e^-\nu_\tau{\bar\nu}_e$ decay. The allowed region of the parameters is given by the relation
$$2(kE-\bm k\cdot\bm q)\leq M^2 \,.$$
 Integrating over the angles of  vector $\bm q$, we find
\begin{align}\label{dae2}
	&dA_{e}=\dfrac{B_{e}\,\mbox{Im}\,b\,d\bm k}{ \pi Eq M^6(1+M^2/2E^2)}\Bigg<4q^3 \left[M^2+2kE (\bm\Lambda\cdot\bm n_{\bm k})^2-6kE\right]\theta(k_0-k)
	\nonumber\\
	&+\Bigg\{\dfrac{1}{2}(\bm\Lambda\cdot\bm n_{\bm k})^2\Bigg[k(E-q)^3(E+3q)
	+\dfrac{M^6}{2k^2}\Bigg(\dfrac{3M^2}{8k}-E\Bigg)\Bigg]
		\nonumber\\
	&-\dfrac{3}{2}k(E-q)^3(E+3q)+M^2(E-q)^2(E+2q)-\dfrac{M^8}{32k^3}\Bigg\}\theta(k-k_0)\Bigg>\,,
\end{align}
 Here $\theta(x)$ is the Heaviside step function, $k_0=(E-q)/2$ and $0\leq k\leq k_{max}$, where $k_{max}=(E+q)/2$.
Thus, the contribution of $\mbox{Re}\,b$ vanishes.  After integration over angles of  vector $\bm k$ the asymmetry reads
 \begin{align}\label{dae3}
 	&dA_{e}=\dfrac{4B_{e}\,\mbox{Im}\,b\,k^2\,dk}{Eq M^6(1+M^2/2E^2)}\Bigg\{4q^3 \left(M^2-\dfrac{16}{3}kE\right)\theta(k_0-k)
 	\nonumber\\
 	&+\Bigg[M^2(E-q)^2(E+2q)-\dfrac{4}{3}k(E-q)^3(E+3q)
 	-\dfrac{EM^6}{12k^2}
 \Bigg]\theta(k-k_0)\Bigg\}\,,
 \end{align}
The dependence of asymmetry $dA_e/dk$ on $k$   is shown in Fig. \ref{asek} for a few values of  energy $E$. It is seen that this dependence is very nontrivial.

The integration of $dA_e/dk$ over $k$ in all allowed region     $0\leq k\leq k_{max}$  gives zero. Therefore, it is naturally to define the total asymmetry $A_e$ as the integral over $k$ in the region $k_{max}/2\leq k\leq k_{max}$. We have
 \begin{align}\label{ae}
	&A_{e}=\dfrac{B_{e}\,\mbox{Im}\,b\,(E+q)}{ 192 Eq (1+M^2/2E^2)}\,
		\left[(11q-3E)\,\theta(E-E_0)+\dfrac{16 q^4 (E+q)^3}{M^6}\,\theta(E_0-E)\right]\,,
\end{align}
where $E_0=3M/\sqrt{8}$. The energy dependence of $A_e$ is shown in Fig. \ref{atot}.

 If one takes in Eq.\eqref{dae1} first the integral over k in the region $0\leq k\leq M^2/[2(E-\bm n_{\bm k}\cdot \bm q)]$ and then over $d\Omega_{\bm q}$, we obtain a simple result    
 \begin{align}\label{daeangle}
 	&dA_{e}=\dfrac{B_{e}\,\mbox{Im}\,b\,d\Omega_{\bm k}}{ 16\pi (1+M^2/2E^2)}\,
 	[3(\bm n_{\bm k}\cdot\bm\Lambda)^2-1]
 	\left[\dfrac{M^2}{2Eq}\ln\left(\dfrac{E+q}{E-q}\right)-1\right]\,.
  \end{align}

 \section{$e^+e^-\to \tau^+\tau^-\to \pi^+\pi^-\nu_\tau{\bar\nu}_\tau$}
 If each $\tau$ lepton decays into a pion and a neutrino (antineutrino), both the imaginary and real parts of $b$ can be measured even in the case of an unpolarized electron beam. It is this case that we consider in this section.  We define the  asymmetry as
 \begin{align}\label{apipi}
 	&dA_{\pi\pi}=\dfrac{d\sigma_{\pi\pi}(\bm k_1,\,\bm k_2)-d\sigma_{\pi\pi}(-\bm k_2,\,-\bm k_1)}{2\sigma_0}\,,
 \end{align}
 where $d\sigma_{\pi\pi}(\bm k_1,\,\bm k_2)$ is the cross section of the process $e^+e^-\to \pi^+\pi^-\nu_\tau{\bar\nu}_\tau$, $\bm k_1$ and $\bm k_2$ are the momenta of $\pi^-$ and $\pi^+$, respectively. For $dA_{\pi\pi}$ we obtain
 \begin{align}\label{dapipi}
 	&dA_{\pi\pi}=\dfrac{3B^2_{\pi}\,d\bm k_1\,d\bm k_2}{ (2\pi)^3M^4(1+M^2/2E^2)\omega_{1}\omega_{2}(E-\omega_{1})(E-\omega_{2})}\nonumber\\
 	&\times\,\int\,d\Omega_{\bm q}\,\delta(E-\omega_{1}-|\bm q-\bm k_1|)\,\,\delta(E-\omega_{2}-|\bm q+\bm k_2|)\nonumber\\
 	&\times\Bigg<\mbox{Re}\,b\,\Bigg\{(\bm q\cdot\bm\Lambda)\,\Big([\bm q\times\bm\Lambda]\cdot(\omega_{2}\bm k_1+\omega_{1}\bm k_2)\Big)\nonumber\\
 	&-\dfrac{\Big(M^2+(\bm q\cdot\bm\Lambda)^2\Big)}{E}\,\Big(\bm q\cdot[\bm k_1\times\bm k_2]\Big)
 	+E(\bm q\cdot\bm\Lambda)\big(\bm \Lambda\cdot[\bm k_1\times\bm k_2]\Big)\Bigg\}\nonumber\\
 	&+
 	\mbox{Im}\,b\,\dfrac{M^2}{2}\Bigg\{\dfrac{(\omega_{2}-\omega_{1})}{E}[M^2+(\bm q\cdot\bm\Lambda)^2]+(\bm q\cdot\bm\Lambda)\big(\bm \Lambda,\bm k_1+\bm k_2\Big)\Bigg\}\Bigg>\,.
 \end{align}
 Integrating over the angles of  vector $\bm q$, we find
 \begin{align}\label{dapipi1}
 	&dA_{\pi\pi}=\dfrac{3B^2_{\pi}\,d\bm k_1\,d\bm k_2}{ (2\pi)^3M^4(1+M^2/2E^2)\omega^2_{1}\omega^2_{2}\,\sqrt{(1-x^2)\,{ q}^2(q^2-{ P}^2)}}\nonumber\\
 	&\times\Bigg<\mbox{Re}\,b\,\Big(\bm\Lambda\cdot[\bm k_1\times\bm k_2]\Big)\Bigg[2(q^2-P^2)(\bm N_2\cdot\bm\Lambda)\nonumber\\
 	&+\dfrac{(\bm P\cdot\bm\Lambda)}{E}\Big(M^2+2P^2-q^2-\dfrac{Ea_2}{1-x}\Big)\Bigg]+
 	\mbox{Im}\,b\,\dfrac{M^2}{2}\Bigg[\dfrac{(\omega_{2}-\omega_{1})}{E}
 	\Big(M^2\nonumber\\
 	&+(\bm N_3\cdot\bm\Lambda)^2(q^2-{ P}^2)+(\bm P\cdot\bm\Lambda)^2\Big)+(\bm P\cdot\bm\Lambda)\big(\bm \Lambda,\bm k_1+\bm k_2\Big)\Bigg]\Bigg>\,,
 \end{align}
 Here the following notation is introduced:
 \begin{align}\label{not}
 	&\bm N_1=\frac{\bm n_1+\bm n_2}{2(1+x)}\,,\quad  \bm N_2=\frac{\bm n_1-\bm n_2}{2(1-x)}\,,\quad \bm N_3=\frac{[\bm n_2\times\bm n_1]}{\sqrt{1-x^2}}\,,\nonumber\\
 	&\bm P=a_1\,\bm N_1+a_2\,\bm N_2\,,\quad a_1=\dfrac{M^2(\omega_{1}-\omega_{2})}{2\omega_{1}\omega_{2}}\,,\quad a_2=2E-\dfrac{M^2(\omega_{1}+\omega_{2})}{2\omega_{1}\omega_{2}}\,,\nonumber\\
 	&x=(\bm n_1\cdot\bm n_2)\,,\quad \bm n_1=\dfrac{\bm k_1}{\omega_{1}}\,,\quad \bm n_2=\dfrac{\bm k_2}{\omega_{2}}\,.
 \end{align}
Note that  the coefficient in front of $\mbox{Im}\,b$  changes its sign at the replacement $\bm n_1\leftrightarrow \bm n_2\,,\, \omega_1 \leftrightarrow\omega_2$, while  the coefficient  in front of $\mbox{Re}\,b$ does not change its sign. 

 It is also interesting to consider the asymmetry that remains after the integration over the angles of  vectors $\bm k_1$ and $\bm k_2$.
 The corresponding result reads
 \begin{align}\label{dapipiEE}
 	&dA_{\pi\pi}=\dfrac{2B^2_{\pi}\,\mbox{Im}\,b\,(\omega_2-\omega_1)d\omega_1\,d\omega_2}{ Eq^2(1+M^2/2E^2)}\,.
 \end{align}
 Here only the imaginary part of $b$ contributes. Integrating over $\omega_2$ in the region $|2\omega_2-E|<q$, we obtain the result
 \begin{align}\label{dapipiE}
 	&dA_{\pi\pi}=\dfrac{B^2_{\pi}\,\mbox{Im}\,b\,(1-2\omega_1/E)d\omega_1}{ q(1+M^2/2E^2)}\,,
 \end{align}
 which is consistent with \eqref{da1}. Integrating \eqref{dapipiE} over $\omega_2$ in the range $(E-q)/2<\omega_2 <  E/2$, we get
 \begin{align}\label{apipiE}
 	&A_{\pi\pi}=\dfrac{B_\pi^2\,\mbox{Im}\,b}{ 4(1+M^2/2E^2)}\,\sqrt{1-\dfrac{M^2}{E^2}}\,.
 \end{align}
 Naturally, $A_{\pi\pi}={B_\pi}A_\pi$.
 
\begin{figure}[h]
 	\centering
 	\includegraphics[width=0.47\linewidth]{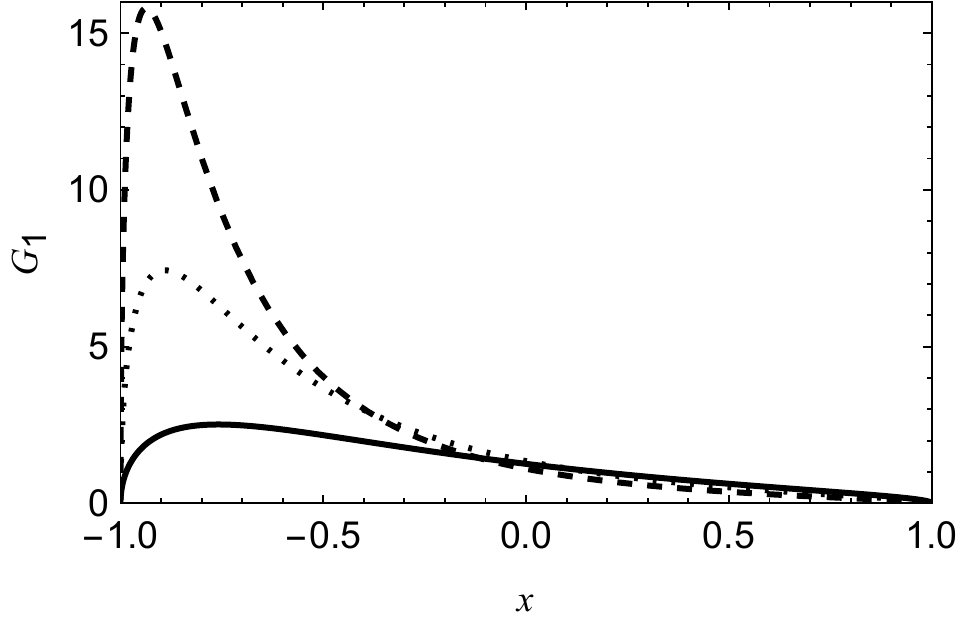}
 	\includegraphics[width=0.47\linewidth]{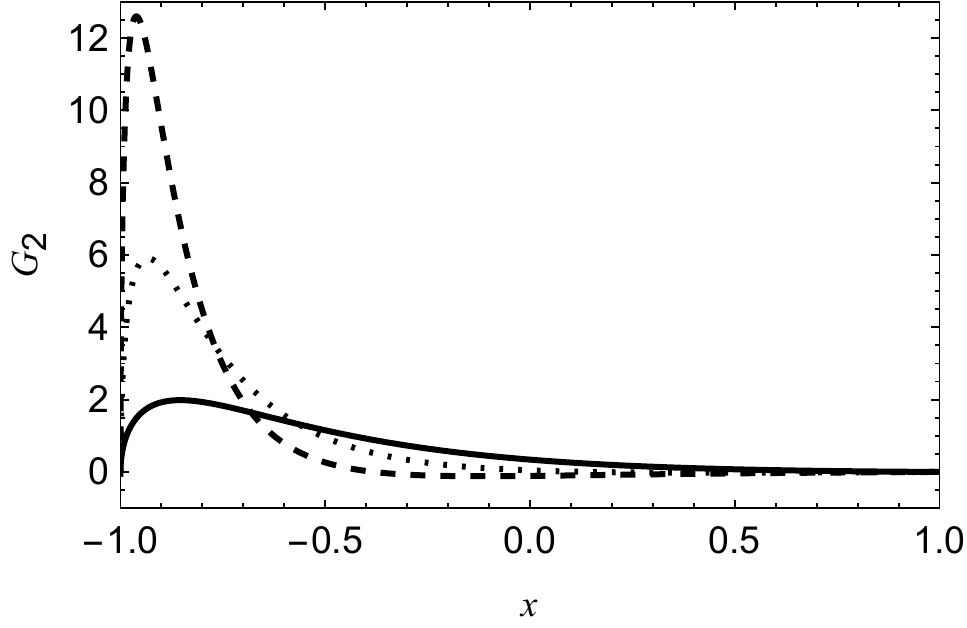}
 	\caption{The dependence of functions $G_1$ (left) and $G_2$ (right) on $x=\bm n_1\cdot\bm n_2$, see \eqref{G12} for $E=1.5\,M$ (solid curve), $E=2\,M$ ( dotted curve) and  $E=2.5\,M$ (dashed curve).}
 	\label{g12}
 \end{figure}

 From our point of view, the most convenient for measurement is the asymmetry integrated over the energies of  emitted pions. Taking first in \eqref{dapipi} the integral over $\omega_1$ and $\omega_2$ and then over $d\Omega_{\bm q}$, we find
 
 \begin{align}\label{dapipiangles}
 	&dA_{\pi\pi}=\dfrac{q^2\,B^2_{\pi}\,d\Omega_1\,d\Omega_2}{(32\pi)^2M^4\,(1+M^2/2E^2)a^2(1+a)^4}\,\Big\{ G^{(1)}_{\pi\pi}[(\bm\Lambda\cdot\bm n_1)^2-(\bm\Lambda\cdot\bm n_2)^2]\,\mbox{Im}\,b\,\nonumber\\
 	&+ G^{(2)}_{\pi\pi}[(\bm\Lambda\cdot\bm n_1)-(\bm\Lambda\cdot\bm n_2)]([\bm n_1\times\bm n_2]\cdot\bm\Lambda)\,\mbox{Re}\,b\Big\}\,,\nonumber\\
 	&G^{(1)}_{\pi\pi}=3(1+a)\,\Bigg\{\Big[a(4 a^2+16 a-3)\,E^2-(a+1)(4a^2+4a+3)q^2\Big]\nonumber\\
 	&+\dfrac{3\ln(\sqrt{a}+\sqrt{1+a})}{\sqrt{a(1+a)}}\Big[a(6 a+1)\,E^2+(a+1)(2a+1)q^2\Big]\Bigg\}\,,\nonumber\\
 	&G^{(2)}_{\pi\pi}=-\dfrac{3}{4}\,\Bigg\{\Big[a(8 a^2-94a+3)\,E^2+(a+1)(8a^2-10a-3)q^2\Big]\nonumber\\
 	&+\dfrac{3\ln(\sqrt{a}+\sqrt{1+a})}{\sqrt{a(1+a)}}\Big[a(24 a^2-12 a-1)\,E^2+(a+1)(8 a^2+4a+1)q^2\Big]\Bigg\}\,,\nonumber\\
 	&a=\dfrac{q^2}{2M^2}[1+(\bm n_1\cdot\bm n_2)]\,.
 \end{align}
 It is seen that the coefficient in front of $\mbox{Im}\,b$  changes its sign at the replacement $\bm n_1\leftrightarrow \bm n_2$ , in contrast to the coefficient  in front of $\mbox{Re}\,b$. This circumstance makes it easier to separate the contributions of  $\mbox{Im}\,b$ and $\mbox{Re}\,b$ to the asymmetry. The dependence of the functions, 
\begin{equation}\label{G12}
	 { G}_{1}= \dfrac{2q^2\sqrt{1-x^2} \, G^{(1)}_{\pi\pi}}{64M^4\,(1+M^2/2E^2)a^2(1+a)^4}\,,\quad  { G}_{2}= \dfrac{q^2\sqrt{2(1-x)(1-x^2)} \, G^{(2)}_{\pi\pi}}{64M^4\,(1+M^2/2E^2)a^2(1+a)^4}\,, 
 \end{equation}
 on $x=\bm n_1\cdot\bm n_2$ is shown in Fig. \ref{g12} for a few values of the energy $E$.

 \section{$e^+e^-\to \tau^+\tau^-\to e^+e^-\nu_\tau{\bar\nu}_\tau\nu_e\bar\nu_e$}
Similar to the asymmetry $dA_{\pi\pi}$, it is possible to measure the asymmetry $dA_{ee}$ in the cross section  $d\sigma_{ee}$ of the process  $e^+e^-\to \tau^+\tau^-\to e^+e^-\nu_\tau{\bar\nu}_\tau\nu_e\bar\nu_e$,
\begin{align}\label{aee}
	&dA_{ee}=\dfrac{d\sigma_{ee}(\bm k_1,\,\bm k_2)-d\sigma_{ee}(-\bm k_2,\,-\bm k_1)}{2\sigma_0}\,,
\end{align}
where $\bm k_1$ and $\bm k_2$ are the momenta of electron and positron, respectively. 
 Again, both $\mbox{Re}\,b$ and $\mbox{Im}\,b$ can be extracted from this asymmetry without using of initial electron polarization. The most convenient from the experimental point of view is the asymmetry in the angular distribution. The straightforward calculations give
\begin{align}\label{daeeangles}
	&dA_{ee}=\dfrac{q^2\,B^2_{e}\,d\Omega_1\,d\Omega_2}{(32\pi)^2M^4\,(1+M^2/2E^2)a^2(1+a)^4}\,\Big\{ G^{(1)}_{ee}[(\bm\Lambda\cdot\bm n_1)^2-(\bm\Lambda\cdot\bm n_2)^2]\,\mbox{Im}\,b\,\nonumber\\
	&+ G^{(2)}_{ee}[(\bm\Lambda\cdot\bm n_1)-(\bm\Lambda\cdot\bm n_2)]([\bm n_1\times\bm n_2]\cdot\bm\Lambda)\,\mbox{Re}\,b\Big\}\,,\nonumber\\
	&G^{(1)}_{ee} =-\dfrac{1}{3}G^{(1)}_{\pi\pi}\,,\quad 
	G^{(2)}_{ee} =\dfrac{1}{9}G^{(2)}_{\pi\pi}\,.
\end{align}
Here $G^{(1)}_{\pi\pi}$ and $G^{(1)}_{\pi\pi}$ are given in Eq.\eqref{dapipiangles}.
Neglecting the muon mass compared to $M$, we obtain the same result \eqref{daeeangles} for asymmetry in the cross section of the processes  $e^+e^-\to \mu^+\mu^-\nu_\tau{\bar\nu}_\tau\nu_\mu\bar\nu_\mu$, $e^+e^-\to \mu^+e^-\nu_\tau{\bar\nu}_\tau\nu_\mu\bar\nu_e$, and $e^+e^-\to \mu^-e^+\nu_\tau{\bar\nu}_\tau\nu_e\bar\nu_\mu$.

\section{Discussion of the results}
In our work we have obtained the asymmetries which contain both $\mbox{Re}b$ and $\mbox{Im}\,b$. 
\begin{figure}[h]
	\centering
	\includegraphics[width=0.55\linewidth]{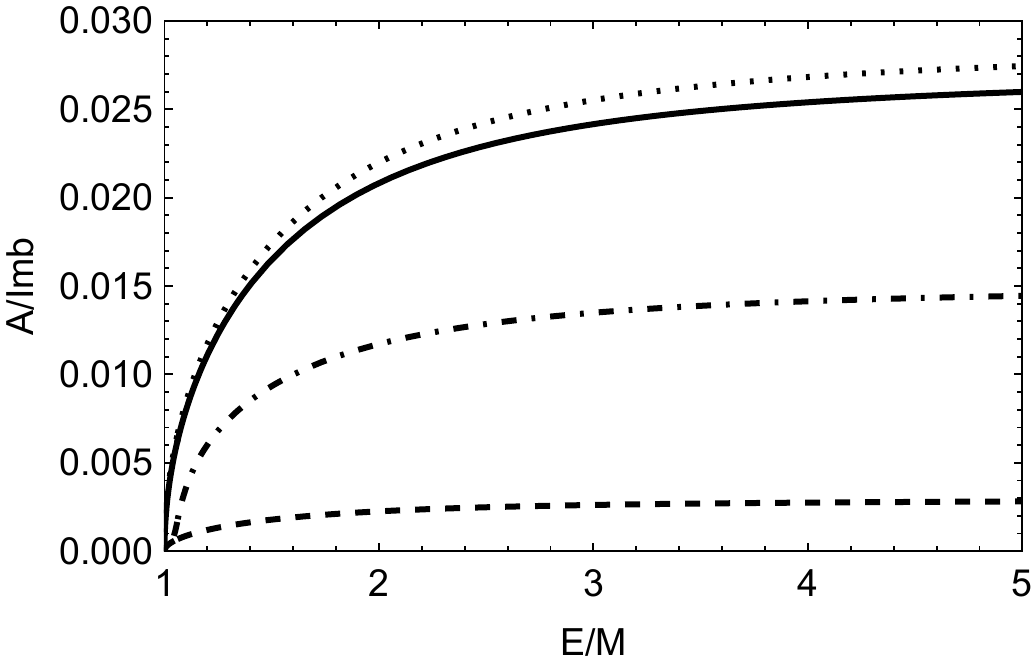}
	\caption{Total asymmetry $A$ in units of $\mbox{Im}\,b$ as a function of energy $E$. Solid curve: $A_\pi$, dotted curve: $A_\rho$,  dash-dotted curve: $A_{e}$, dashed curve: $A_{\pi\pi}$.}
	\label{atot}
\end{figure}
Asymmetries \eqref{api},\eqref{arho}, \eqref{dae}, \eqref{apipi}, and \eqref{aee}  are the odd quantities  with respect to $\mbox{CP}$ transformations. Indeed, as a result of this transformation
\begin{align*}
	&d\sigma_\pi^{(-)}(\bm k)\rightarrow \,d\sigma_\pi^{(+)}(-\bm k)\,,\quad d\sigma_\pi^{(+)}(-\bm k)\rightarrow\, d\sigma_\pi^{(-)}(\bm k)\,,\\
		&d\sigma_\rho^{(-)}(\bm p,\,\bm f)\rightarrow\, d\sigma_\rho^{(+)}(-\bm p,\,-\bm f)\,,\quad d\sigma_\rho^{(+)}(-\bm p,\,-\bm f)\rightarrow\, d\sigma_\rho^{(-)}(\bm p,\,\bm f)\,,\\
			&d\sigma_e^{(-)}(\bm k)\rightarrow \,d\sigma_e^{(+)}(-\bm k)\,,\quad d\sigma_e^{(+)}(-\bm k)\rightarrow\, d\sigma_e^{(-)}(\bm k)\,,\\
	&d\sigma_{\pi\pi}(\bm k_1,\,\bm k_2)\rightarrow\, d\sigma_{\pi\pi}(-\bm k_2,\,-\bm k_1)\,,\quad d\sigma_{\pi\pi}(-\bm k_2,\,-\bm k_1)\rightarrow\, d\sigma_{\pi\pi}(\bm k_1,\,\bm k_2)\,,\nonumber\\
		&d\sigma_{ee}(\bm k_1,\,\bm k_2)\rightarrow\, d\sigma_{ee}(-\bm k_2,\,-\bm k_1)\,,\quad d\sigma_{ee}(-\bm k_2,\,-\bm k_1)\rightarrow\, d\sigma_{ee}(\bm k_1,\,\bm k_2)\,.
\end{align*}
The term $\propto\gamma^\mu$ in \eqref{tauvert} is $\mbox{CP}$-even while the term 
 $\propto\,b$  is  $\mbox{CP}$-odd. Therefore, asymmetries appear due to interference between $\mbox{CP}$-odd and  $\mbox{CP}$-even terms and are linear in $b$.   
The energy dependence of total asymmetries is shown in Fig. \ref{atot}. The solid curve corresponds to $A_\pi$, the dotted curve to $A_\rho$, the dash-dotted curve to $A_{e}$, and the dashed curve to $A_{\pi\pi}$. The  curves corresponding to $A_\pi$, $A_\rho$, and $A_{\pi\pi}$ have the same energy dependence and differ only in scale. Indeed, it follows from Eqs.~ \eqref{apiE}, \eqref{arhoE}, \eqref{apipiE}  that
$$A_\rho=1.06\,A_{\pi}\,,\quad A_{\pi\pi}=0.108\,A_{\pi}\,.$$
Though the formulas  for  $A_e$ and $A_\pi$  are completely different, the  corresponding curves  have   similar shapes.

\section{Conclusion} 
In conclusion, we have considered the processes $e^+e^-\rightarrow \tau^+\pi^-\nu_\tau$, $e^+e^-\rightarrow \pi^+\tau^-\bar\nu_ \tau$, $e^+e^-\rightarrow \tau^+\rho^-\nu_\tau$,$e^+e^-\rightarrow \rho^+\tau^-\bar\nu_\tau $, $e^+e^-\to \tau^+e^-\nu_\tau{\bar\nu}_e\,,\,$ and $\,e^+e^-\to \tau^-e^+\nu_e{\bar\nu}_\tau$  with longitudinally polarized electrons, as well as the processes $e^+e^-\rightarrow \pi^+\pi^-\nu_\tau\bar\nu_\tau$, $e^+e^-\to e^+e^-\nu_\tau{\bar\nu}_\tau\nu_e\bar\nu_e$, $e^+e^-\to \mu^+\mu^-\nu_\tau{\bar\nu}_\tau\nu_\mu\bar\nu_\mu$, $e^+e^-\to \mu^+e^-\nu_\tau{\bar\nu}_\tau\nu_\mu\bar\nu_e$, and $e^+e^-\to \mu^-e^+\nu_\tau{\bar\nu}_\tau\nu_e\bar\nu_\mu$ with unpolarized electrons for the invariant masses $\sqrt {s}\ll m_Z$ of the initial electron and positron.  We have  calculated analytically the $\mbox{CP}$-odd  asymmetries $\propto\mbox{Re}\,b$ and $\propto\mbox{Im}\,b$. Measuring these quantities can improve the upper limits for $\mbox{Re}\, b$ and $\mbox{Im}\, b$. It is shown that to measure $\mbox{Im}\,b$, polarization is not needed, and to measure $\mbox{Re}\,b$, the polarization is not necessary, but simplifies the measurement.

\section*{Acknowledgements}
We are grateful to D.A. Epifanov and V.S. Vorobiev for valuable discussions.

\end{document}